\documentclass[11pt,a4paper,twoside]{article}
\usepackage{changepage} % to use adjustwidth
\usepackage{graphicx}
\usepackage{verbatim}
\usepackage{gensymb}
\usepackage{amsmath,amssymb,amsfonts} % Typical maths resource packages
\usepackage{bm}

\usepackage[usenames,dvipsnames]{color}
\usepackage[table]{xcolor}
\definecolor{light-gray}{gray}{0.90}

\usepackage[T1]{fontenc}
\usepackage{lmodern}
\usepackage{titlesec}
\titleformat{\section}{\normalfont\Large\color{black}}{\bf\thesection}{1em}{}

\titleformat{\subsection}{\normalfont\large\color{black}}{\bf\thesubsection}{1em}{}

\usepackage[a4paper,inner = 2.5cm, outer = 3.0cm, top = 2.5cm, bottom = 2.5cm]{geometry} %changes the inner, outer, top and bottom margins
\usepackage{fancyhdr}
\fancypagestyle{plain}{ %
  \fancyhf{} % remove everything
}

\usepackage{epstopdf}
\DeclareGraphicsExtensions{.png,.jpg,.jpeg,.eps}
\graphicspath{ {./Figures/} }
\usepackage{float}
\usepackage{caption}
\usepackage{subcaption}
\usepackage{booktabs}

\usepackage[square,comma,numbers,sort&compress]{natbib} % Use author style or numbers, also sort and compress the citation and put like [1-4]
\usepackage{url} 
\usepackage[colorlinks]{hyperref}
\hypersetup{linkcolor=blue, citecolor = blue}
\usepackage{siunitx}
\usepackage{authblk}

\title{On-chip quantum interference with heralded photons from two independent micro-ring resonator sources in silicon photonics}

\author[]{Imad I. Faruque\thanks{imad.faruque@bristol.ac.uk}}
\author[]{Gary F. Sinclair}
\author[]{Damien Bonneau}
\author[]{John G. Rarity}
\author[]{Mark G. Thompson}
\affil[]{Quantum Engineering Technology Labs, H. H. Wills Physics Laboratory \& Department of Electrical and Electronic Engineering, University of Bristol, Merchant Venturers Building, Woodland Road, Bristol BS8 1UB, United Kingdom.}
\date{}                     %% if you don't need date to appear
\begin{document}
\maketitle
%\Addresses
\begin{abstract}
High visibility on-chip quantum interference among indistinguishable single-photons from multiples sources is a key prerequisite for integrated linear optical quantum computing. Resonant enhancement in micro-ring resonators naturally enables brighter, purer and more indistinguishable single-photon production without any tight spectral filtering. The indistinguishability of heralded single-photons from multiple micro-ring resonators has not been measured in any photonic platform. Here, we report on-chip indistinguishability measurements of heralded single-photons generated from independent micro-ring resonators by using an on-chip Mach-Zehnder interferometer and spectral demultiplexer. We measured the raw heralded two-photon interference fringe visibility as $72\pm3\%$. This result agrees with our model, which includes device imperfections, spectral impurity and multi-pair emissions. We identify multi-pair emissions as the main factor limiting the nonclassical interference visibility, and show a route towards achieving near unity visibility in future experiments.
\end{abstract}
%\vspace{10pt}

\section*{Introduction}
Integrated photonics represents a promising concept for quantum photonic technologies. Several platforms have achieved notable success constructing the fundamental building blocks for on-chip quantum computing. However, we are still at an early stage with regards to the full-scale integration of optical and electronic components into a single monolithic \cite{Gentry2015, Sun2015, PhysRevLett.118.100503}, heterogeneous \cite{komljenovic2016} or hybrid architecture \cite{Meany2014}. Among the key challenges remaining is the integration of several identical high-performance single-photon sources \cite{Varnava2008}. To be considered ideal, a source should: emit single-photons on demand (deterministic), produce a high rate of single-photons (bright) and emit each photon in a single mode (pure). In addition, it should be possible to construct many such identical sources, such that the photons they produce are indistinguishable. A wide variety of single-photon sources have been demonstrated across several material platforms that exhibit, or promise, some degree of on-chip integration. These sources can be broadly characterised as single-emitter systems (quantum dots, colour centres, etc) and photon-pair sources based on parametric nonlinearities. Recent advances in quantum dot sources have demonstrated single-photon production with high spectral purity ($>92\%$) for resonant excitation and moderately high extraction efficiency \cite{somaschi2015near, Ding2016}. However, achieving both metrics simultaneously remains highly challenging \cite{iles2017phonon}, particularly in a waveguide integrated form \cite{LPOR:LPOR201500321}. In addition, the ability to produce several such indistinguishable sources remains unproven \cite{PhysRevB.89.035313, PhysRevLett.111.237403}. Also most single-emitters require cryogenic temperature (except some colour centres such as \cite{Marshall2011}). Nonetheless, quantum dot sources have found recent application as a high-brightness source for boson sampling \cite{PhysRevLett.118.130503,wang2017high}, allowing much faster collection of scattering statistics than has been achieved using parametric sources. 

In contrast, parametric photon sources have proven themselves readily integrable, operate at room temperature and have demonstrated high purity and indistinguishability in a wide range of two-photon experiments \cite{Silverstone2013, Silverstone2015, Jin2014, PhysRevLett.118.100503, wang2017experimental,PhysRevApplied.4.021001}. Four-photon heralded Hong-Ou-Mandel (HOM) interference \cite{PhysRevLett.96.240502} has also been demonstrated between independent sources with a visibility of above $90\%$ for directly laser written silica waveguides \cite{Spring2017a}, $88\%$ in silicon-on-insulator (SOI) \cite{Zhang2016} and microstructured optical fibre \cite{Fulconis2007} and $93\%$ in Periodically-Poled Lithium Niobate (PPLN) sources \cite{Aboussouan2010}, although none of these four-photon experiments have yet been performed in a fully-integrated form. Parametric sources are however, intrinsically non-deterministic and must be operated at low brightness to avoid the deleterious effects of higher-order photon number terms. Nonetheless, this can be mitigated by the active multiplexing of several such sources \cite{Migdall2002, Jeffrey2004, Meany2014, Collins2013, Xiong2016a}.

All the four-photon heralded Hong-Ou-Mandel (HOM) interference experiments \cite{Harada2011, Zhang2016} in SOI have been restricted to the use of non-resonant sources (waveguides). However, in our work we employ narrow-linewidth ring resonators for photon-pair generation. This allows for the more efficient and compact generation of photon-pairs \cite{Savanier:16,doi:10.1063/1.4711253} (compared to linear waveguide sources) and eliminates the requirement for tight spectral filtering of the generated photon-pairs to improve the purity of the heralded photons \cite{Vernon2015a, Vernon:16}, as is often required in linear waveguide sources. 

In this work, we demonstrate for the first time the on-chip quantum interference between single-photons heralded from two independent micro-ring resonators. Also, photon-pair generation, spectral demultiplexing and nonclassical interference are all fully-integrated onto a monolithic silicon photonic chip, representing an important step in the on-chip integration of quantum optical experiments. 

Numerous previous works have demonstrated the combined on-chip generation and manipulation of single-photons from resonators, but have been restricted to the generation of a single photon-pair \cite{PhysRevApplied.4.021001}. In contrast, our work shows the on-chip interference among photons heralded from two independent sources as 4-fold coincidences, allowing us to estimate the indistinguishability among multiple micro-ring resonators. This implies the feasibility of on-chip integration of two or more micro-ring resonators as heralded single-photon sources, as they all have to be identical for scalable integration for photonic quantum computing. A 4-fold measurement also allows us to directly explore the effect that photon purity, indistinguishability and the presence of higher photon-number contamination have on the all-important visibility of nonclassical interference. 

\begin{figure}[htbp]
\centering
\includegraphics[width=\linewidth]{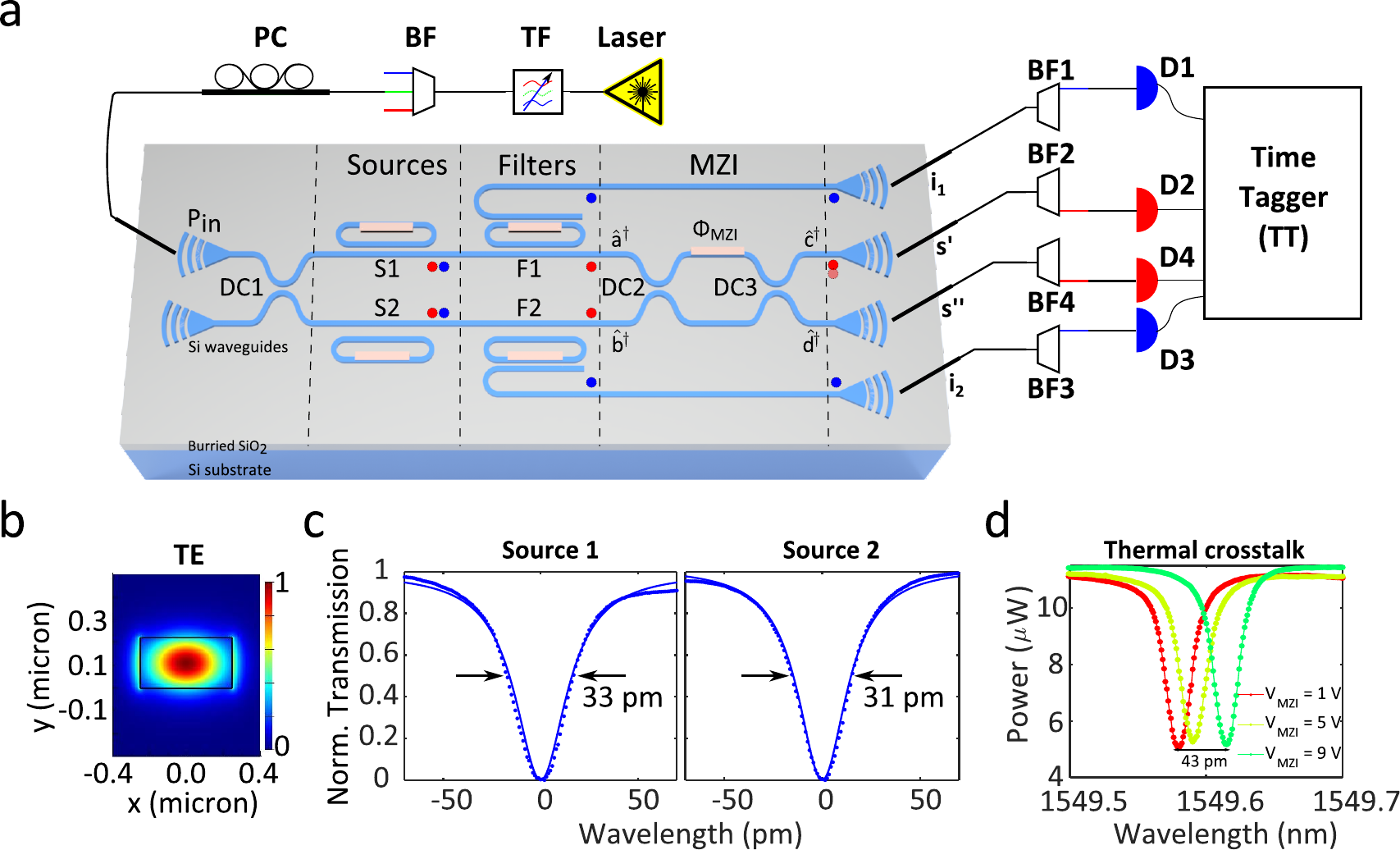}
\caption{\label{fig:chip} (a) shows the experimental setup. A pulsed laser passes through a broadband filter (BF) and a tunable filter (TF) to suppress broadband background emission and to match the bandwidth of the laser to the micro-ring resonator sources. A polarisation controller (PC) is used to optimise transmission onto the chip via a grating coupler. The photonic circuit consists of a directional coupler (DC1), micro-ring resonator sources (S1 and S2), micro-ring resonator filters (F1, F2), and an MZI composed of directional couplers (DC2, DC3) and a thermal phase-shifter ($\Phi_\mathrm{MZI}$). Photon-pairs generated by the sources are coupled off-chip and filtered (BF1 to BF4) and collected by single-photon detectors (D1 to D4) connected to a time-tagger (TT). Analysis of four-fold coincidences is done in postprocessing. (b) The electric field intensity of the fundamental transverse electric (TE) mode of the silicon nanowire waveguide used in the chip is calculated with \textit{Lumerical Mode Solver}. Waveguide dimensions were \SI{500}{nm} $\times$ \SI{220}{nm} and the group index was estimated as $n_g=4.16$ at the wavelength \SI{1550}{nm}. (c) Spectral profiles of the source micro-ring resonators S1 and S2. Both sources are seen to be largely spectrally indistinguishable, with linewidths of \SI{33}{pm} and \SI{31}{pm} (Q-factor $\sim5\times10^4$). (d) Due the sensitivity of the high Q-factor sources, the resonance shifts when the heater of the MZI dissipates heat. In this figure, the resonance of the source S2 shifts \SI{43}{pm}, when the MZI dissipates \SI{60}{mW} heat ($2\pi$ phase) from its off position. Counteracting this thermal crosstalk is essential to perform the experiment and explained in detail in Appendix A.}
\end{figure}

\section*{Experimental Setup}
Our experimental setup is illustrated in Fig.~\ref{fig:chip}(a). It consists of three major parts: laser pulse preparation for photon-pair generation; the reconfigurable photonic circuit for on-chip single-photon generation and indistinguishability measurements and single-photon detection system with logic unit. The chip is fabricated by \textit{Institute of Microelectronics} Singapore through a standard multi-project wafer run.

We chose a \SI{50}{MHz} repetition rate pulsed laser (Pritel, FFL) as a pump for spontaneous four wave mixing photon-pair generation. The laser emits secant hyperbolic solitonic pulses which are nearly transform limited. The central wavelength of the laser is tuned to Channel 39 (\SI{1546.12}{nm}) of International Telegraph Union (ITU) frequency grid. A tuneable bandwidth filter (Yenista XTA-50, TF in Fig.~\ref{fig:chip}(a)) is used to match the bandwidth of the pump laser to the linewidth of the micro-ring resonators used as sources ($\sim$\SI{30}{pm}, Fig.~\ref{fig:chip}(c)). The bandwidth was set to \SI{200}{pm} to suppress the generation of background photons (from spurious four wave mixing) in the input coupling waveguides (Appendix A), while at the same time, filling the entire pump resonance of the sources, allowing for a maximum heralded photon purity \cite{Helt2010}. The background laser noise is suppressed by the use of two consecutive \SI{200}{GHz} channel spacing Dense Wavelength Division Multiplexers (DWDM) as denoted by BF in Fig.~\ref{fig:chip}(a). BFs reduce the background laser noise by at least \SI{80}{dB} at the wavelengths where signal-idler photon-pairs will be generated. Afterwards, an optimal polarisation is chosen using the polarisation controller (PC) to couple the transverse electric (TE) mode into the chip. This is because the waveguide width (\SI{500}{nm}), and the vertical grating couplers are optimised for TE polarisation mode.

The vertical grating couplers (VGCs) couple the light in (and out of) the chip with an average efficiency of $35.5\%$ (\SI{-4.5}{dB} loss). At this point, approximately \SI{1}{mW} of power is in the chip. After the VGC, the pump is split by a directional coupler (DC1) and routed towards the micro-ring resonator sources (S1 and S2). Photon-pairs, historically called signals and idlers, are produced in S1 and S2 by a nonlinear parametric process: spontaneous four wave mixing (SFWM). Both S1 and S2 are thermally tuned to be resonant with the pump (Channel 39) and have a free-spectral range such that signal-idler photon-pairs are produced in Channel 31 (\SI{1552.52}{nm}) and Channel 47 (\SI{1539.77}{nm}) of the ITU grid. Micro-ring resonator add-drop filters (F1) and (F2) are also thermally tuned to be resonant with the idler photons and route these off-chip for heralding using detectors D1 and D3. The heralded signal photons are passed on to the integrated Mach-Zehnder Interferometer (MZI) section of the chip. The MZI is formed of two directional couplers (DC2, DC3) and a thermal phase-shifter ($\Phi_\mathrm{MZI}$). After the MZI interference, the signal photons are coupled off-chip, filtered to suppress residual pump and coupled to single-photon detectors (D2 and D4). Just before the detectors, the aforementioned suppression of the residual pump is performed using \SI{200}{GHz} DWDM filters on all four signal-idler channels. We note that this filtering is solely to suppress pump noise, and is much wider than the narrow bandwidth idler photons generated by the ring resonator sources ($\sim$ \SI{3.8}{GHz}), so does not have an appreciable influence on the heralded photon purity. After pump suppression, the photons are detected using commercially available Superconducting Nanowire Single Photon Detectors (SNSPDs - Photon Spot, Inc.). All four detectors (D1 to D4: average efficiency 75\%, dark counts $<$ \SI{200} {counts/s}) are connected to a picosecond-resolution timetagger (TT - HydraHarp, PicoQuant). The timetagger records the photon arrival times which are post-processed to identify four-fold coincidence events ($P_\mathrm{4f}$) as a function of the MZI phase ($\Phi_\mathrm{MZI}$). $P_\mathrm{4f} (\Phi_\mathrm{MZI})$ represents the interference fringe. 
\begin{figure}[htbp]
\centering
\includegraphics[width= \linewidth]{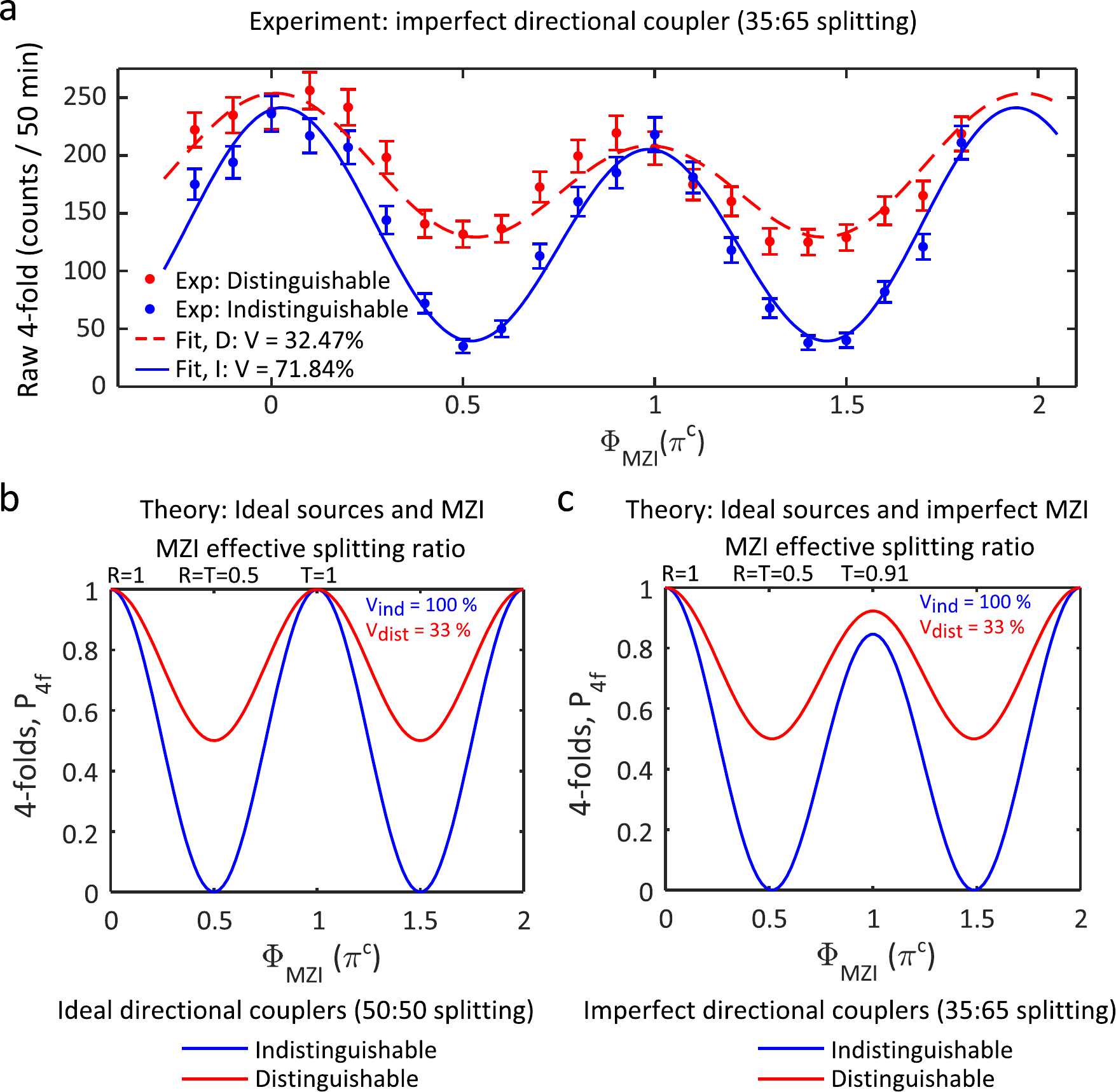}
\caption{\label{fig:fringes}4-fold coincidences as a function of the MZI phase. (a) The experimentally measured 4-fold coincidence fringes, demonstrating a fitted visibility of 72\% for indistinguishable photons. The reduced visibility compared to the ideal case arises due to several factors which are included in the fit: contributions from higher photon-number terms; spectral impurity of the heralded photons and potentially the spectral distinguishability of the sources. These are discussed in the \textit{Results and Analysis} section. (b) Theoretical 4-fold coincidence probability assuming ideal sources and MZI. The MZI fringe visibility (defined by Eq.~(\ref{eqn:Vmzi})) has a maximum of 100\% for indistinguishable photons and 33\% for distinguishable photons. When the phase, $\phi_\mathrm{MZI}$, is adjusted to $\pi/2$ the MZI has a 50:50 splitting ratio ($R=T=0.5$, where $R$ and $T$ are the reflection and transmission coefficients of the MZI). (c) Theoretical 4-fold coincidence probability assuming ideal sources, but an imperfect MZI. The MZI is assumed to be constructed from two non-ideal directional couplers (see DC2 and DC3 of Fig.~\ref{fig:chip}) that have a 35:65 splitting ratio, as was the case for our fabricated chip. It is seen that the shape of the fringe changes as a result of the imperfect splitting ratio of the directional couplers. However, the global maximum and minimum probabilities achieved for both distinguishable and indistinguishable fringes remains the same. The main effect of the non-balanced splitting ratio of the directional couplers is to reduce the effective transmitivity of the MZI at the fringe peak where $\phi_\mathrm{MZI}=\pi$ (T$<$ 1). This does not reduce the fringe peak at $\phi_\mathrm{MZI}=0$ (R=1) and so will not limit the fringe visibility. The details of the theoretical models and the full fittings that includes spectral impurity and multi-pair (upto 10 photon-pairs) emissions are in Appendix B.}
\end{figure}

One of the experimental considerations was to mitigate the thermal crosstalk between the photonic components. In this reconfigurable circuit, when the large heater of the MZI is used to change the phase, the dissipated heat shifts the resonance position of the resonator sources S1 and S2 as much as \SI{43}{pm} and \SI{31}{pm} respectively: this is termed as thermal crosstalk. A full parametric model of the heat compensation has been developed to keep all the four resonators (S1, S2, F1, F2) at the same resonance wavelength for any phase configuration during the experiment, as explained in Appendix A.

\section*{Results and Analysis}\label{sec:results}
The performance of our heralded single-photon sources is reflected in the the measured visibility of the four-fold coincidence fringe, as shown in Fig.~\ref{fig:fringes}(a). This fringe shows the rate of four-fold coincidence events as a function of the MZI phase ($\Phi_\mathrm{MZI}$). Typically, the visibility of a MZI fringe \cite{PhysRevLett.65.1348} is defined by,
\begin{equation}
\label{eqn:Vmzi}
V_\mathrm{MZI} = \frac{(P_\mathrm{4f})_\mathrm{max} - (P_\mathrm{4f})_\mathrm{min}}{(P_\mathrm{4f})_\mathrm{max} + (P_\mathrm{4f})_\mathrm{min}}.
\end{equation}
As shown in Figs.~\ref{fig:fringes}(b) and \ref{fig:fringes}(c) above, for ideal sources producing pure single-photons, we expect a fringe visibility of 100\% for indistinguishable photons and 33\% for distinguishable photons. Here, the imperfect directional couplers (DC2, DC3) of the MZI does not influence the value of the visibility, in contrast to a Hong-Ou-Mandel experiment \cite{Hong1987}. The deviation of the splitting ratio of the directional couplers from its ideal value (balanced 50:50) is due to fabrication imperfections. The splitting ratio of our directional couplers were found to be 35:65. This resulted in a distortion of the measured fringe shape, but did not impact on the maximum visibility observed, as discussed in Appendix B. Briefly, the distortion occurs due to incomplete destructive interferences of some specific MZI phase values for the imperfect directional couplers. 

In our experiment, a fringe visibility of $V_\mathrm{MZI}=71.84\pm3.1\%$ (95\% confidence interval) was found for indistinguishable photons, and $V_\mathrm{MZI}=32.47\pm3.0\%$ for distinguishable photons by fitting the data using the model described by Eq.~(\ref{eq:4F_nov}) of the Appendix B, that takes into account the factors discussed below. There are several factors that can lead to a reduction in the fringe visibility, such as spectral distinguishability between the sources, reduced spectral purity of the heralded single-photons and the contribution of higher photon-number terms.
\begin{figure}[htbp]
\centering
\includegraphics[width= \linewidth]{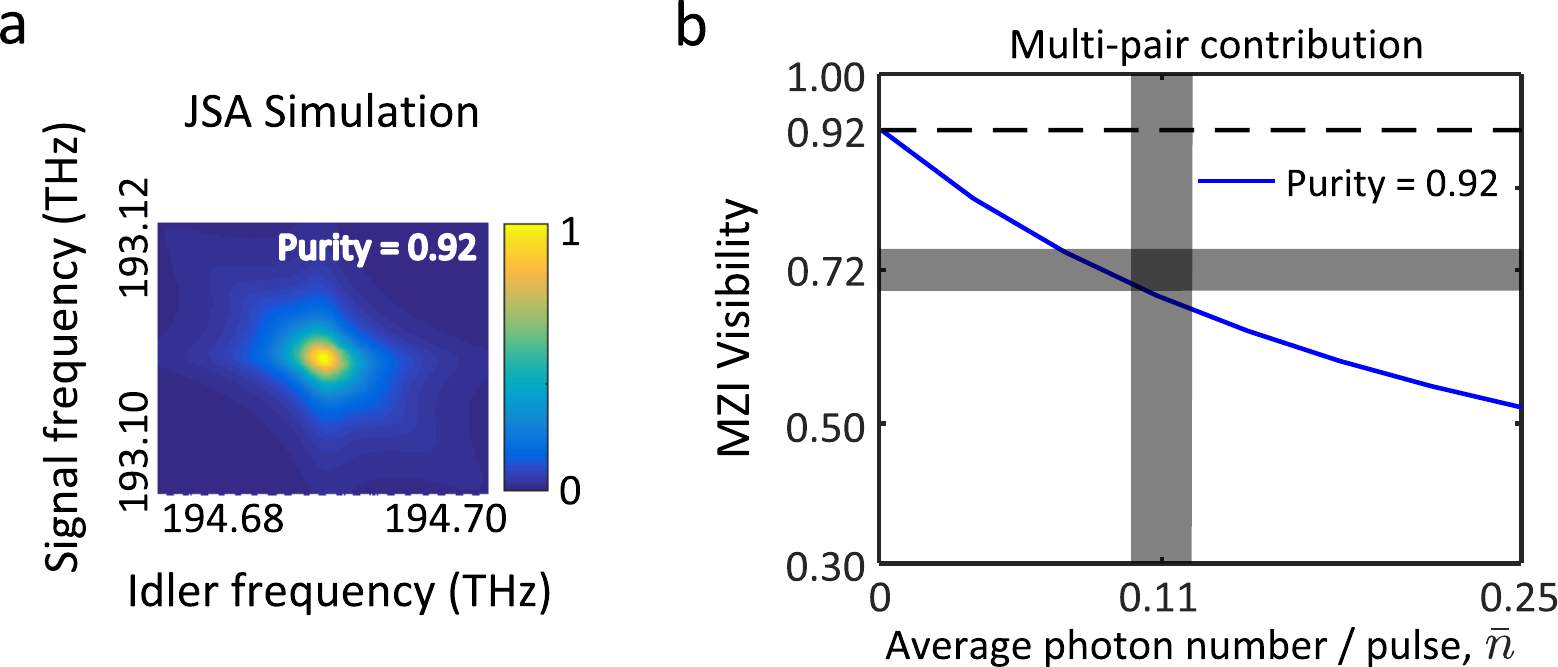}
\caption{\label{fig:SpectraAndVisibility}(a) Simulation of the joint spectral amplitude (JSA) of the micro-ring resonators used for this experiment shows a spectral purity of $92\%$. (b) Visibility of the MZI fringe as a function of the source brightness, $\bar{n}$ including multi-mode and multi-pair emissions. Identical sources with a purity of $0.92$ are considered which which can be modelled well by two effective Schmidt modes. A total 10 photon-pairs generated from both of the sources at a time are considered to include the multi-pair effects. In our experiment we determined an average source brightness of $0.110\pm0.012$ photons produced from each source per pump pulse. From the experiment, the MZI fringe visibility is found as $72\pm3\%$. The simulated graph (blue solid line) goes through the intersection of these error margins, showing the agreement between the simulation and the experiment.}
\end{figure}

We begin by considering the spectral distinguishability of both ring resonator sources. As shown in Fig.~\ref{fig:chip}(c), having a Full Width Half-Maximum (FWHM) linewidth of \SI{33}{pm} and \SI{31}{pm} for sources S1 and S2 respectively, both sources had an excellent spectral overlap. In addition, as discussed in the Appendix A, by careful control of the chip temperature and consideration of the thermal crosstalk between components, the stability of the overlap was ensured. We do not therefore expect the spectral distinguishability of the sources to make a significant contribution to the reduced fringe visibility. We note that the spectral purity of photons heralded from a ring resonator source has been shown to have an upper limit of $93\%$, by calculating purity from the joint spectral amplitude (JSA) \cite{Helt2010}. Following the same procedure as in \cite{Helt2010}, the JSA simulation of our sources S1 and S2 designs, with DWDM filtering shows $92\%$ purity as shown in Fig.~\ref{fig:SpectraAndVisibility}(a). Therefore, our sources are seen to be approaching the optimal purity for a simple ring resonator design. It has been proposed that by tailoring the ratio of the linewidths at the pump and photon-pair wavelengths, it should be possible to achieve arbitrarily high purities from a modified ring design \cite{Vernon2017}. We should not therefore view the current non-ideal purity as a fundamental limit of all resonant sources, but rather solely a limitation of our current basic micro-ring resonator design.

The final contribution to the reduced fringe visibility comes from the contribution of multiphoton terms. Due to the probabilistic nature of our parametric photon-pair sources, we know that sometimes two or more photons-pairs will be produced simultaneously from a single source. Without photon-number resolving detectors, we are unable to distinguish the heralding of these higher photon-number states from the single-photons that we wish to interfere in the MZI stage of the chip. Several authors have developed models describing the effect of higher photon-number terms on the visibility of nonclassical interference \cite{Fulconis2007, Harada2011, Spring2017a}. A full model considering the effect of these higher photon-number terms, along with the impurity of the heralded photons (approximated by the first two Schmidt modes), is developed in the Appendix B. The effect of the pump brightness (which leads to higher photon-number terms) on the MZI fringe visibility is shown in Fig.~\ref{fig:SpectraAndVisibility}(b) for impure ($\sim92\%$ purity) heralded photonic states. The brightness of a photon-pair source is often represented by the number of photon-pairs generated per pump pulse, $\bar{n}$. The solid blue line shows that the MZI fringe visibility reduces with $\bar{n}$ due to multi-pair contamination. Our micro-ring resonator sources have an average brightness of $\bar{n} = 0.110\pm0.012$ within 95\% confidence interval. As shown in the figure, within the margin of errors of the measured visibility and the average brightness (represented by semi-transparent grey rectangles), the modelled visibility matches. Therefore, the model agrees with our experimental results and identifies multi-pair contamination as the most drastic cause of visibility reduction. It also shows that if the multi-photon contamination is removed (e.g. using photon-number resolving detectors), we can achieve $92\%$ interference visibility.

To determine the brightness of our ring sources, we examine the two-fold coincidence counts as a function of optical input power (Appendix C). From this we determine a brightness of $\bar{n}_\mathrm{S1} = 0.093\pm0.001$ and $\bar{n}_\mathrm{S2} = 0.123\pm0.001$ with 68\% confidence interval for each source. Ideally, we would reduce the contribution of higher photon-number terms by reducing the pump power and collecting four-fold coincidence events over a longer time period. However, the four-fold count rate is currently limited by system losses, largely due to the grating couplers ($\sim$ \SI{4.5}{dB} loss) and off-chip filtering ($\sim$ \SI{2}{dB}) and channel loss before the detection. In future, the use of higher efficiency grating couplers \cite{Ding:14} ($\sim$ \SI{0.6}{dB}) and on-chip filtering \cite{Piekarek:17} both have the potential to greatly improve the system detection efficiency. This should allow the use of lower pair production rates at the source and therefore reducing higher photon-number contamination. 

\section*{Conclusions}
Our experiment demonstrates a high degree of integration with sources, spectral demultiplexing add-drop filters and an MZI all on a monolithic silicon chip and successfully independently controlled by thermal tuning. Using this chip, we have interfered two heralded single-photons from two independent micro-ring resonators for the first time, enabling a MZI raw fringe visibility of 72\%. In contrast to previous experiments, which generally interfered a single path-entangled photon-pair, our experiment allows for the direct evaluation of the indistinguishability of independently generated heralded photons - a vital prerequisite for the scalable construction of all optical quantum computers.

We also identify the main factors that are currently limiting the non-unity fringe visibility: residual impurity of the heralded single-photons and the contribution of higher photon-number terms to the fringe visibility. As discussed above, a clear path exists to overcoming both of these constraints. Designs have recently been proposed that should allow near unity heralded photon purity by suitable engineering of the ring resonator coupling at pump, signal and idler wavelengths \cite{Vernon2017}. Higher photon-number terms can be removed by using photon number resolving detectors and/or improving the system detection efficiency. Improving the system detection efficiency of our single-photon detectors, primarily by reducing grating coupler losses, will allow weaker pumping of the micro-ring resonator sources whilst maintaining a suitably high four-fold coincidence rate. Operating in this regime, the reduction of fringe visibility to higher photon-number terms is known to be mitigated \cite{Fulconis2007, Harada2011, Spring2017a}. Thus, our fringe visibility of 72\% should be seen as a basis from which higher visibilities can be achieved in future, with some appropriate photonic engineering. 

\section*{Acknowledgements}
This work was supported by the Marie Curie Actions within the Seventh Framework Programme for Research of the European Commission, under the Initial Training Network PICQUE, Grant No. 608062, the Engineering and Physical Sciences Research Council (EPSRC) through grant EP/L024020/1 and The European Research Council (ERC). M.G.T. acknowledges fellowship support from the EPSRC (EP/K033085/1). J.G.R. acknowledge fellowship support from the (EPSRC, UK). The author thanks Dr. Hugo Cable, Dr. Jonathan Mathews for useful discussions and Dr. Philip Sibson for lending the HydraHarp.

\appendix
\begin{figure}[htbp]
\centering
\includegraphics[width= \linewidth]{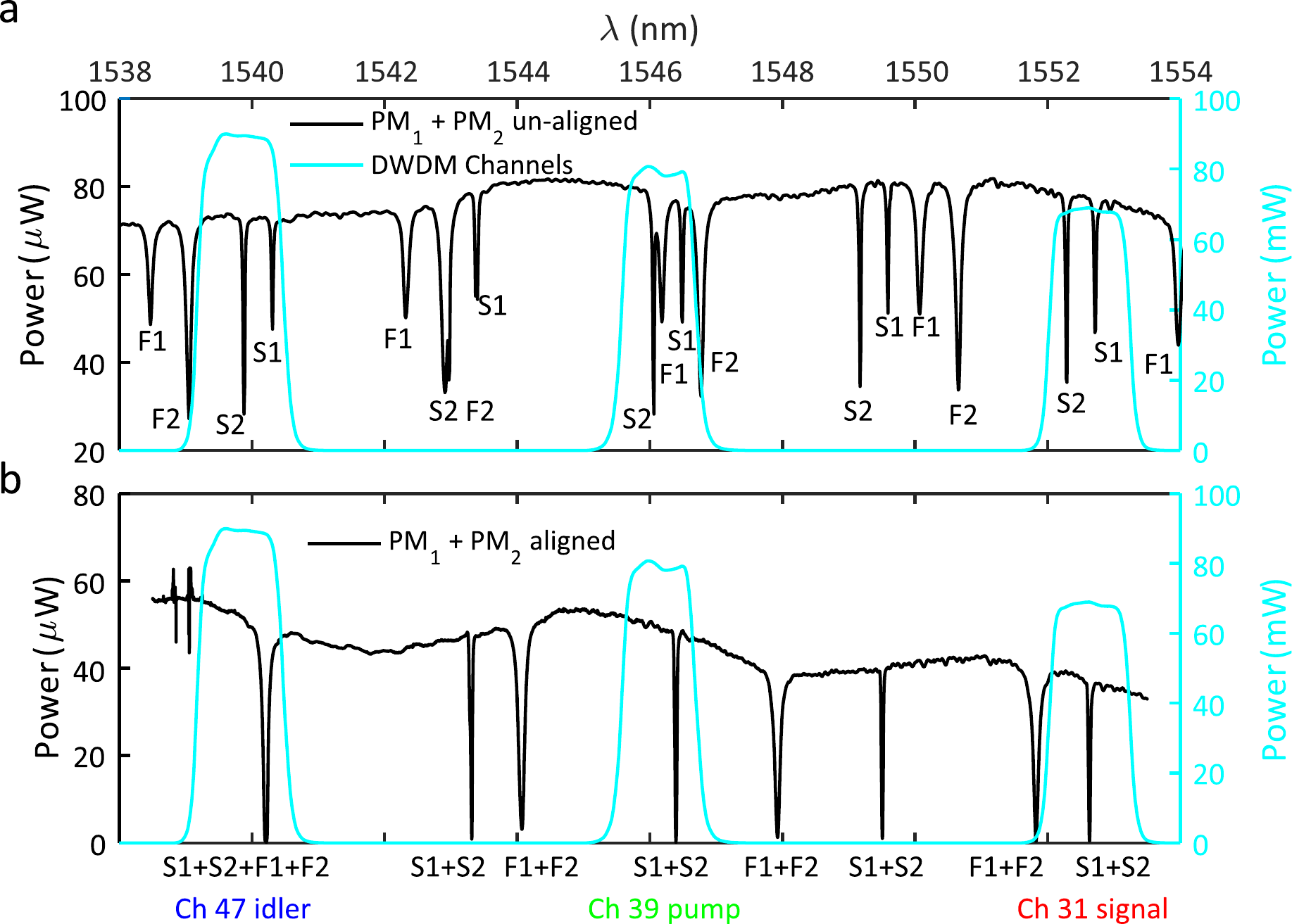}
\caption{\label{fig:classical_characterisation}Spectral response of the device by adding the two output ports of the MZI ($\mathrm{PM}_i$: power-meter value of the $i^{th}$ port). (a) Without any spectral alignments, all the resonators S1, S2, F1, F2 are in spectrally distinguishable. (b) Using the thermal phase-shifter to align S1, S2 to match the the resonances with the DWDMs' ITU grid channels 47, 39 and 31 for idler, pump and signal spectra respectively. }
\end{figure}
\section*{Appendix A: spectral response and thermal crosstalk}
One of the principle challenges of this experiment is to maintain spectral alignment between sources S1 and S2 while the MZI thermal phase-shifter, $\Phi_{MZI}$, is swept over a 2$\pi$ range (approximately 0 to \SI{42}{mW} power). Due to the narrow resonance linewidth of the micro-ring resonator sources (shown in Fig.~\ref{fig:chip}(c)), the micro-ring resonators are sensitive to thermo-optic crosstalk from neighbouring heaters, which are used to align the sources, filters and perform the MZI phase shift. The spectral response of the whole device is shown in Fig.~\ref{fig:classical_characterisation} with and without spectral alignments using the on-chip thermal phase-shifters.

\begin{figure}[htbp]
\centering
\includegraphics[width= \linewidth]{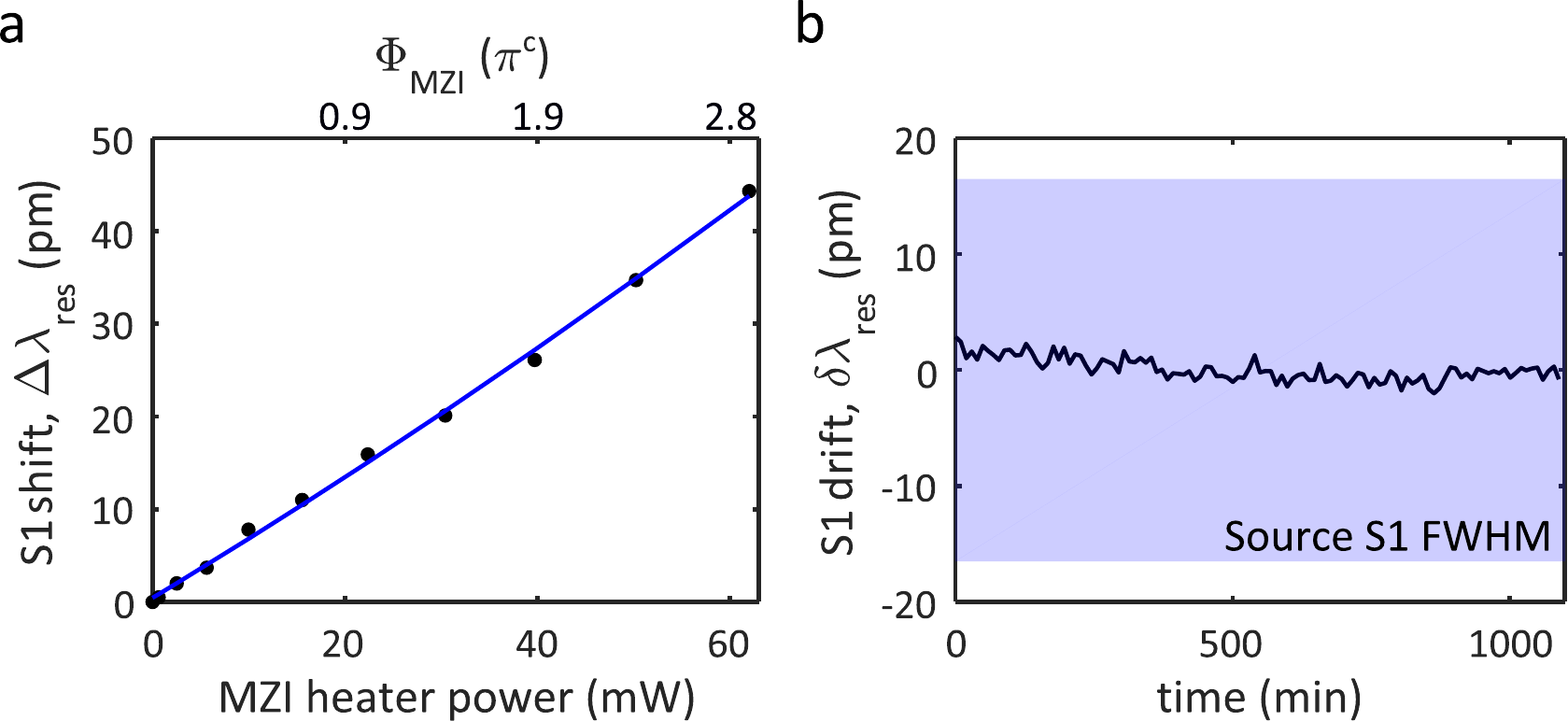}
\caption{\label{fig:thermal}Wavelength stability of micro-ring resonator sources. (a) Thermal cross-talk between the MZI heater and source S1. In this test, the position of source S1 is measured, while the MZI heater power is adjusted over its full range. A large thermal crosstalk is observed between the MZI heater and source S1, which needs to be compensated for. This is achieved, by reducing the heater power supplied directly to S1 in proportion to the thermal crosstalk from the MZI heater. In this way, the resonance position of S1 can be held constant. The same technique is applied between S2, F1 and F2 and the MZI heater to ensure stability of both sources and on-chip filters. (b) Measured position of the micro-ring resonator source S1 over a period of 1100 minutes, which is equal to the time required for the acquisition of a complete fringe. The shaded area represents the FWHM of the resonance and the black trace represents a series of 111 measured resonance peak positions. The measured resonance positions have a standard deviation of \SI{1.0}{pm}, compared to the FWHM of \SI{33}{pm} for Source 1 (relative drift: 7.1\%).}
\end{figure}

In practice, the MZI heater is the dominant source of thermal crosstalk, due the large range of powers over which it operates. Figure~\ref{fig:thermal}(b) demonstrates the excellent stability of the micro-ring resonator sources whilst in the quiescent state over the time required to take a complete four-photon fringe. The chip temperature is maintained by use of an Arroyo Instruments 5240 Thermoelectric Controller, which ensures a temperature stability better than \SI{0.01}{K} (corresponding to a micro-ring resonator stability of \SI{0.7}{pm}). The micro-ring resonator position was measured by scanning a low-power CW probe (Yenista T100S-HP) to determine the resonance position from the transmission spectrum, whilst calibrating the wavelength against a wavemeter (Bristol Instruments 721B-IR). However, when the MZI heater power is altered, a shift in the resonance position of the sources is observed (see Fig.~\ref{fig:chip}(d), Fig.~\ref{fig:thermal}(a)). This is due to the relatively high power of the MZI heater (up to \SI{60}{mW}) and the quite close proximity of the heater to the micro-ring resonator source (200$\sim$\SI{350}{\micro \meter}). To compensate for thermal crosstalk, a characterisation was first performed to measure the shift in each micro-ring position (S1, S2, F1, F2) as a function of the MZI heater power. Therefore, by reducing the power supplied directly to each micro-ring as the MZI heater power was increased, the resonance position could be maintained whilst performing a fringe. Electrical control of the on-chip heaters was performed using a prototype high-resolution voltage driver with current readout capability (Qontrol Systems LLP).

\begin{figure}[htbp]
\centering
\includegraphics[width= \linewidth]{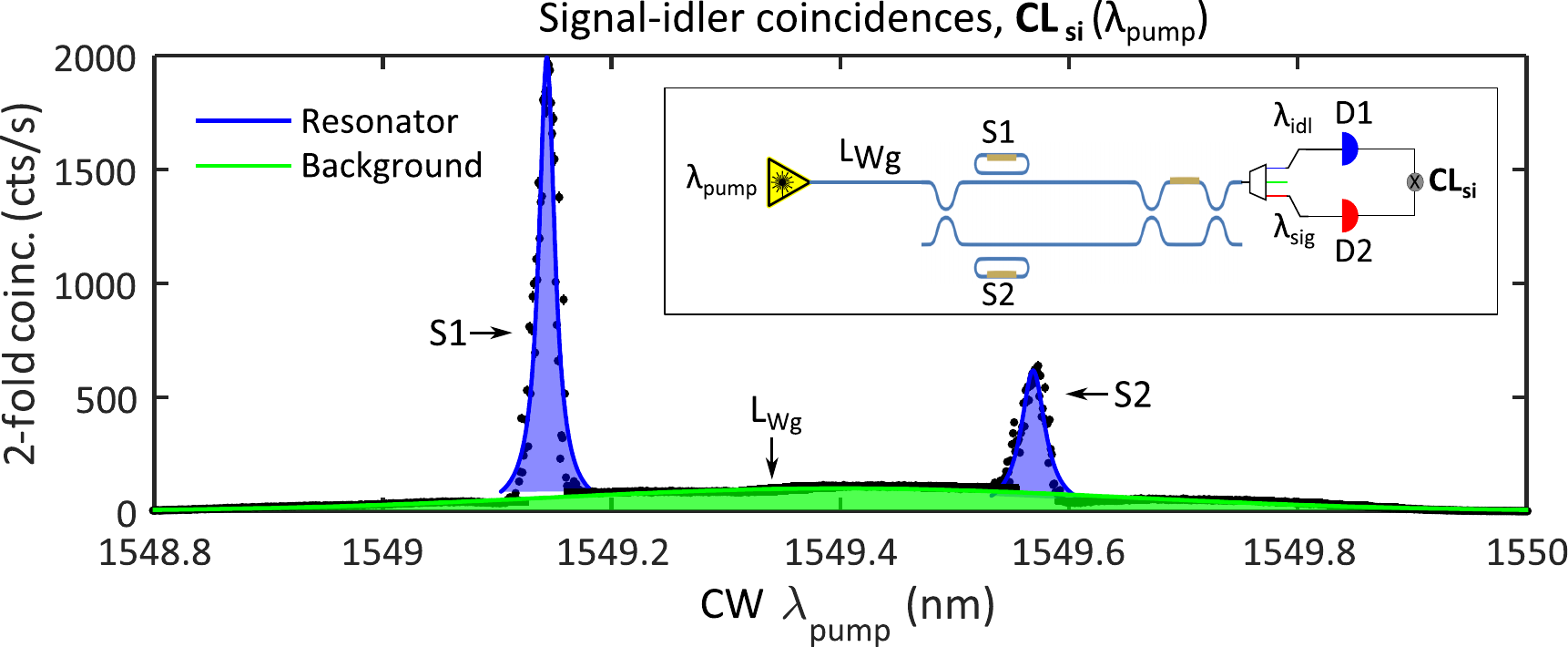}
\caption{\label{fig:Sp_FWM}Spurious four wave mixing in the input waveguide ($L_{WG}\sim$\SI{730}{\micro\meter}) is substantial over the whole bandwidth of the DWDM channels. As shown in the inset, a CW pump laser ($\lambda_p$) is scanned over Ch 39 of the ITU grid, while signal-idler photon-pairs are collected as 2-fold coincidences for each pump wavelength $\lambda_p$ from Ch 47 and Ch 31. The filter resonators F1 and F2 are detuned off the pump channel for this experiment. The plot shows the background photon-pair generated by the waveguide in shaded green, and by the resonators by shaded blue. If the accumulated source and background photons are $C_{S1}$, $C_{S2}$, and $C_{Bg}$, then $(C_{S1}+C_{S2})/C_{Bg}\approx1.05$.}
\end{figure}

When using the full pump-width it was noticed that the input waveguide from the VGC to the DC1 was long enough ($\sim$\SI{730}{\micro\meter}) such that the pump produced noticeable amount of four wave mixing photon-pairs in the waveguide. These pairs are identified as spurious four wave mixing photon-pairs and considered as background contributions. As our detectors are not wavelength sensitive, they accumulate all of the photon-pairs generated in the waveguide and the micro-resonator sources S1 and S2. The total accumulated background photons from the waveguide can be substantial over the whole bandwidth of the DWDM channels. This is because the high-Q micro-resonators generate a significantly higher number of photon-pairs per unit bandwidth, but in a very narrow wavelength range ($\sim$\SI{30}{pm}) compared to the bandwidth of the DWDM channels ($\sim$\SI{1100}{pm}). As shown in Fig.~\ref{fig:Sp_FWM}, if the accumulated photons in the blue shades and green shades are $C_{S1}$, $C_{S2}$, and $C_{Bg}$, then $(C_{S1}+C_{S2})/C_{Bg}\approx1.05$. For a 2-fold coincidence experiment, this value can modify the actual result from the resonators. A tuneable bandwidth filter is used after the laser to narrow down the pump pulse to \SI{200}{pm} FWHM. Considering the solitonic $sech$ pulses are transform limited, the narrowed FWHM will reduce the peak power of the pulses significantly. Thus, the amount of photon-pairs produced in the waveguide will reduce drastically, while the spectral power density coupled into the micro-ring (and hence photon production rate) will remain largely unchanged.

\section*{Appendix B: Theoretical model}\label{sec:theory}
The nonclassical interference of single-photons is an essential requirement for linear-optical quantum computing. Here, we briefly review the nonclassical interference behaviour expected in our four-photon triggered fringe experiment. In the first section, we initially assume our sources, interferometer and detectors are all ideal. In our experiment however, it was found that fabrication imperfections resulted in a splitting ratio of the directional couplers that was 35:65, rather than the ideal 50:50 ratio of the intended design. Therefore, we next examine how deviations from the nominal splitting ratio of the on-chip directional couplers will affect the shape of the four-photon fringe, but will not limit the observed maximum fringe visibility. Afterwards, the effects of various experimental imperfections (multi-mode fields, multi-pair emission) discussed in the \textit{Results and Analysis} section are then included in the model to derive a equation to fit the MZI fringe.
\subsection*{Ideal Sources and MZI}
We begin by supposing that a pair of pure single-photons are simultaneously heralded from both sources (S1 and S2 in Fig.~\ref{fig:chip}). We assume that they are indistinguishable in all degrees of freedom, other than being in separate spatial modes: heralded photons generated in S1 are denoted as being in spatial mode `a' and photons heralded from S2 are in spatial mode `b'. The heralded initial state is therefore $\vert \psi_\mathrm{in} \rangle = \hat{a}^\dag \hat{b}^\dag \vert \mathrm{vac} \rangle$, where $\hat{a}^\dag$ and $\hat{b}^\dag$ are the single-mode creation operators for photons in spatial modes `a' and `b' respectively and $\vert \mathrm{vac} \rangle$ is the vacuum state. In the Heisenberg picture, for an ideal MZI with perfectly balanced directional couplers (50:50 splitting ratio) the creation operators for the input modes (`a' and `b') and output modes (`c' and `d') are related by,
\begin{equation}
\left ( \begin{array}{c} \hat{c}^\dag \\ \hat{d}^\dag \end{array} \right ) = \frac{1}{2} \left ( \begin{array}{cc} 1 & 1 \\ 1 & -1 \end{array} \right ) \left ( \begin{array}{cc} e^{i \phi_\mathrm{MZI}} & 0 \\ 0 & 1 \end{array} \right ) \left ( \begin{array}{cc}1 & 1 \\ 1 & -1 \end{array} \right ) \left ( \begin{array}{c} \hat{a}^\dag \\ \hat{b}^\dag \end{array} \right ),
\end{equation}
where $\phi_\mathrm{MZI}$ is the phase shift applied in one arm of the MZI (see Fig.~\ref{fig:chip}). We note that the transformation effected by the MZI can also be written as the transformation of a variable beam splitter \cite{0034-4885-66-7-203}, with some phase shifts on the input and output modes:
\begin{equation}
\label{eqn:VBS}
\left ( \begin{array}{c} \hat{c}^\dag \\ \hat{d}^\dag \end{array} \right ) = e^{i \phi_\mathrm{MZI}/2} \left ( \begin{array}{cc} 1 & 0 \\ 0 & i \end{array} \right ) \left ( \begin{array}{cc} \cos (\phi_\mathrm{MZI}/2) & \sin (\phi_\mathrm{MZI}/2) \\ \sin (\phi_\mathrm{MZI}/2) & -\cos (\phi_\mathrm{MZI}/2) \end{array} \right ) \left ( \begin{array}{cc} 1 & 0 \\ 0 & i \end{array} \right ) \left ( \begin{array}{c} \hat{a}^\dag \\ \hat{b}^\dag \end{array} \right ).
\end{equation}
 The probability of a four-fold coincidence is given by the overlap of the output state with the state describing the simultaneous scattering of one photon into output mode `c' and the other into output mode `d'. That is, for indistinguishable single-photons,
\begin{equation}
\label{eqn:Pind}
P^\mathrm{ind}_\mathrm{4f}(\phi_\mathrm{MZI}) = \vert \langle \psi_\mathrm{out} \vert \hat{c}^\dag \hat{d}^\dag \vert \mathrm{vac} \rangle \vert^2 = \left \vert \langle \psi_\mathrm{out} \vert (R-T) \hat{a}^\dag \hat{b}^\dag \vert vac \rangle \right \vert^2 = \frac{1}{2} \left [1 + \cos \left (2 \phi_\mathrm{MZI} \right ) \right ],
\end{equation}
where $R=\sin^2(\phi_\mathrm{MZI}/2)$ and $T=\cos^2 (\phi_\mathrm{MZI}/2)$ are the effective reflection and transmission coefficients of the MZI ($R+T=1$). These two terms ($R$ and $T$) arise from the simultaneous reflection or simultaneous transmission of both input photons into the output modes. We note that the phase of these terms always results in a destructive interference between both processes, due to the unitarity of the transformation Eq.~(\ref{eqn:VBS}). A minimum of the four-fold coincidence probability is seen to occur at the point $\phi_\mathrm{MZI}= \pi/2$, where $R=T=0.5$ and therefore $P^\mathrm{ind}_\mathrm{4f}=0$ (see Fig.~\ref{fig:fringes}(a)). That is, this corresponds to the point at which the MZI is perfectly balanced (50:50 splitting ratio) and is exactly the situation in which we expect to observe HOM interference between the indistinguishable photon-pairs.

Over the range $\phi_\mathrm{MZI} \in [0, 2 \pi)$ the four-fold coincidences peak when $\phi_\mathrm{MZI} = \{ 0, \pi \}$. Examining the MZI transformation given by Eq.~(\ref{eqn:VBS}) we see that these correspond to points at which the MZI is completely transmitting or completely reflecting. That is, $\phi_\mathrm{MZI}=0$, and hence $T=\cos^2(\phi_\mathrm{MZI}/2) = 1$ (complete transmission) or $\phi_\mathrm{MZI}=\pi$, with $R=\sin^2 (\phi_\mathrm{MZI}/2)=1$ (complete reflection). In both of these situations, only the simultaneous reflection of both input photons ($R=1$) or simultaneous transmission of both input photons ($T=1$) will contribute a non zero term in Eq.~(\ref{eqn:Pind}). Due to the absence of destructive interference between both of these processes we therefore expect a maximum in the probability of four-fold counts between both independent sources at these two points ($P^\mathrm{ind}_\mathrm{4f}=1$). 

We turn now to the situation in which the heralded photons are distinguishable in some way, for example, when both photons are heralded at different times. Now, the photons are assumed to be generated not only in separate spatial modes, as before, but also in separate temporal modes. We denote the non-overlapping temporal modes by placing a prime on the corresponding mode operators. That is, we now have two sets of input and output mode operators: \{$\hat{a}, \hat{b}, \hat{c}, \hat{d}$\} for photons heralded at time $t=t_1$, and \{$\hat{a}', \hat{b}', \hat{c}', \hat{d}'$\} for photons heralded at time $t=t_2$. Both sets of mode operators are assumed to satisfy the MZI transformation equations given by Eq.~(\ref{eqn:VBS}). In this case, the initial state input to the MZI is given by a pair of photons heralded in distinguishable spatial {\it and} temporal modes $\vert \psi_\mathrm{in} \rangle = \hat{a}^\dag \hat{b'}^\dag \vert \mathrm{vac} \rangle$. Given that we have input two photons in separate temporal modes, the four-fold coincidence probability is now given by the sum of two terms, each one corresponding to the temporally distinguishable photons scattering into either of the two output spatial modes:
\begin{equation}
\label{eqn:Pdist}
P^\mathrm{dist}_\mathrm{4f}(\phi_\mathrm{MZI}) = \vert \langle \psi_\mathrm{out} \vert \hat{c}^\dag \hat{d'}^\dag \vert \mathrm{vac} \rangle \vert^2 + \vert \langle \psi_\mathrm{out} \vert \hat{c'}^\dag \hat{d}^\dag \vert \mathrm{vac} \rangle \vert^2= \frac{1}{2} + \frac{1}{4} \left [1 + \cos \left ( 2 \phi_\mathrm{MZI} \right ) \right ].
\end{equation}
From these equations for the four-fold coincidence probability, Eq.~(\ref{eqn:Pind}) and Eq.~(\ref{eqn:Pdist}), we see that the nonclassical destructive interference of indistinguishable single-photons results in a fringe with much higher contrast than that produced by distinguishable photons. The probabilities for a four-fold coincidence for indistinguishable Eq.~(\ref{eqn:Pind}) and distinguishable photons Eq.~(\ref{eqn:Pdist}) are depicted in Fig.~\ref{fig:fringes}(b). Maxima and minima of the four-fold probability remain at the same positions, but the minimum four-fold probability at $\phi_\mathrm{MZI}=\pi/2$ has increased to $P^\mathrm{dist}_\mathrm{4f}=0.5$, indicating the absence of non-classical HOM interference due to the temporal distinguishability of the heralded photons. In the above, we have compared the interference of distinguishable and indistinguishable single-photons. However, a comparison could also be made between indistinguishable single-photons and thermal states (as a representative `classical' state of the light field). In this case, it has been shown that a maximum HOM visibility of 33\% is found \cite{PhysRevA.60.593, Harada2011} (assuming ideal detectors and low average photon number).

\subsection*{Effect of an Imperfect Splitting Ratio of the MZI Directional Couplers}
In our experiment, it was found that the splitting ratio of the on-chip directional couplers deviated significantly from the intended 50:50 ratio. Although this is seen to result in a change to the shape of the interference fringe, we show that it does not limit the achievable fringe visibility, as shown below. We begin again by assuming a pair of photons heralded in spatial modes `a' and `b', which are otherwise indistinguishable in all degrees of freedom: $\vert \psi_\mathrm{in} \rangle = \hat{a}^\dag \hat{b}^\dag \vert \mathrm{vac} \rangle$, where all symbols have the same meanings as defined above. The input and output modes of the MZI are then related by,
\begin{equation}
\label{eqn:vbsmodes}
\left ( \begin{array}{c} \hat{c}^\dag \\ \hat{d}^\dag \end{array} \right ) = \left ( \begin{array}{cc} \sin \theta & \cos \theta \\ \cos \theta & -\sin \theta \end{array} \right ) \left ( \begin{array}{cc} e^{i \phi_\mathrm{MZI}} & 0 \\ 0 & 1 \end{array} \right ) \left ( \begin{array}{cc}\sin \theta & \cos \theta \\ \cos \theta & -\sin \theta \end{array} \right ) \left ( \begin{array}{c} \hat{a}^\dag \\ \hat{b}^\dag \end{array} \right ),
\end{equation}
where the splitting ratio ($\eta_{DC}$) of the directional couplers (DC2 and DC3) are assumed to be equal and are determined by $\theta$, such that $\eta_{DC} = \sqrt{sin(\theta)}$, and the relative phase shift imparted on one arm of the MZI is given by $\phi_\mathrm{MZI}$. Again, this relationship between the input and output modes of the MZI can be expressed as a variable beamsplitter, with phase shifts on the input and output modes \cite{0034-4885-66-7-203}:
\begin{equation}
\left ( \begin{array}{c} \hat{c}^\dag \\ \hat{d}^\dag \end{array} \right ) = e^{i \phi_\mathrm{MZI}/2} \left ( \begin{array}{cc} e^{i \Phi/2} & 0 \\ 0 & e^{-i \Phi/2} \end{array} \right ) \left ( \begin{array}{cc} -i \cos (\eta/2) & -\sin (\eta/2) \\ \sin ( \eta/2) & i \cos (\eta/2) \end{array} \right ) \left ( \begin{array}{cc} i e^{i \Phi / 2} & 0 \\ 0 &  -i e^{-i \Phi/2} \end{array} \right ) \left ( \begin{array}{c} \hat{a}^\dag \\ \hat{b}^\dag \end{array} \right ).
\end{equation}
Here, the effective splitting ratio of the MZI is determined by $\eta= 2 \arcsin \left [ \sin (\phi_\mathrm{MZI}/2) \sin (2 \theta) \right ]$ and the input and output mode phase shifts are given by $\displaystyle \Phi = -\frac{i}{2} \log \left [ \frac{1 + e^{i \phi_\mathrm{MZI}} \tan^2 \theta}{\tan^2 \theta + e^{i \phi_\mathrm{MZI}}} \right ]$. A four-fold coincidence occurs when heralded single-photons from each source are simultaneously detected on output modes `c' and `d'. Using Eq.~(\ref{eqn:vbsmodes}) to relate the input and output modes, we find that the four-fold detection probability for indistinguishable single-photons is,
\begin{equation}
\label{eqn:PindImperfect}
P^\mathrm{ind}_\mathrm{4f} (\phi_\mathrm{MZI}) = \left \vert \langle \psi_\mathrm{out} \vert \hat{c}^\dag \hat{d}^\dag \vert \mathrm{vac} \rangle \right \vert^2 = \left \vert \langle \psi_\mathrm{out} \vert (R-T) \hat{a}^\dag \hat{b}^\dag \vert \mathrm{vac} \rangle \right \vert^2 =\left [ 2\sin^2 (\phi_\mathrm{MZI}/2) \sin^2 (2 \theta) - 1\right ]^2,
\end{equation} 
where,
\begin{align}
	& R=\sin^2(\eta/2)=\sin^2(\phi_\mathrm{MZI}/2) \sin^2(2 \theta) \\
	& T=\cos^2 (\eta/2)=\left \vert \cos (\phi_\mathrm{MZI}/2) - i \sin(\phi_\mathrm{MZI}/2) ( \sin^2 \theta-\cos^2 \theta ) \right \vert^2
\end{align}
are the effective reflection and transmission coefficients for the MZI. In our experiment, characterisation of the chip using classical light showed that the splitting ratio of the directional couplers was 35:65, which corresponds to a value of $\theta=\arccos (\sqrt{0.65}) = 0.63$ in Eq.~(\ref{eqn:vbsmodes}). The four-fold coincidence probability assuming this imperfect splitting ratio is plotted in Fig.~\ref{fig:fringes}(c). We see that the imperfect splitting ratio of the directional couplers results in a slight change in shape of the fringe, particularly in the region around $\phi_\mathrm{MZI}=\pi$. At this point, a local maximum occurs in the four-fold probability, with $P^\mathrm{ind}_\mathrm{4f}(\phi_\mathrm{MZI}=0) = \cos^2 (4 \theta) \le 1$. However, whereas $\phi_\mathrm{MZI}=\pi$ previously corresponded to complete transmission ($T=1$), in the MZI with imperfect directional couplers we have $T=\cos^2 (2 \theta) < 1$. In this case, the MZI no longer acts as a perfect cross-over between the input and output modes ($T<1, R \ne 0)$, and there will be a residual destructive interference between both terms in Eq.~(\ref{eqn:PindImperfect}) that give rise to coincidences. This results in a reduction of the coincidence probability ($P^\mathrm{ind}_\mathrm{4f}<1$) at $\phi_\mathrm{MZI}=\pi$. Nonetheless, this does not preclude our ability to reach a maximum of the four-fold coincidence probability, since another maximum occurs at the point $\phi_\mathrm{MZI}=0$, at which point the MZI is completely reflecting to the input photons ($R=1$). This point of complete reflection can be achieved irrespective of the directional coupler splitting ratio, and ensures that we can measure at least one point where the coincidence probability is a maximum.

We note that the imperfect splitting ratio of the directional couplers also has the effect of slightly changing the position of the four-fold minima given by Eq.~(\ref{eqn:PindImperfect}). By finding the stationary points of Eq.~(\ref{eqn:PindImperfect}) we see that the minima occur where $\cos (\phi_\mathrm{MZI})=\csc^2(2 \theta)+ 1$. Nonetheless, these points continue to correspond to a balanced effective splitting ratio of the MZI ($R=T=0.5$) and therefore remain the point of maximum nonclassical interference between the heralded photon-pairs ($P^\mathrm{ind}_\mathrm{4f}=0$). Thus, although we have shown that the fringe shape does change due to the imperfect splitting ratio of the constituent directional couplers, we nonetheless see that the global maximum and minimum points on the four-fold probability fringe are not changed.

For completeness, we also derive the probability of a four-fold coincidence event, given a pair of distinguishable photons interfering in an imperfect MZI. As before, we say that a pair of heralded single-photons are generated by sources S1 and S2 in separate spatial modes, which we denote `a' and `b', and in separate temporal modes, which we distinguish by placing a prime next to one of the mode labels: thus, $\vert \psi_\mathrm{in} \rangle = \hat{a}^\dag \hat{b'}^\dag \vert vac \rangle$. Using the transformation for the mode operators in the Heisenberg picture Eq.~(\ref{eqn:vbsmodes}) we find that the four-fold coincidence probability for a pair of distinguishable photons can be written as:
\begin{equation}
\label{eqn:PdistImperfect}
P^\mathrm{dist}_\mathrm{4f} = \left \vert \sin^2(\phi_{MZI}/2)\sin^2(2 \theta) \right \vert^2 + \left \vert 1 - \sin^2(\phi_{MZI}/2)\sin^2(2 \theta) \right \vert^2.
\end{equation}
Again, it can be shown that maxima and minima of the four-fold probability occur at the same values of $\phi_\mathrm{MZI}$ as for the indistinguishable photon fringe given above Eq.~(\ref{eqn:Pind}). Examination of the global maxima and minima of Eq.~(\ref{eqn:PdistImperfect}) similarly shows that despite the change in fringe shape due to the imperfect splitting ratio of the directional couplers, the global maximum and minimum fringe values remain unchanged.

In summary, although the fringe shape is modified due to the imperfect splitting ratio of the directional couplers in our chip, we nonetheless find that the global minimum and maximum of the four-fold probability fringes remain unchanged. This ensures that the individual fringe visibilities for both distinguishable and indistinguishable photon-pairs are not reduced due to this device imperfection. Importantly, the HOM visibility (discussed below) that depends on the global minimum of both fringes will also remain unchanged as a result of the fringe distortion. Essentially this is due to our continued ability to achieve a 50:50 effective splitting ratio of the MZI, despite the fabrication imperfections of the directional couplers.

\subsection*{MZI fringe, $P_\mathrm{4f}$, including impurity and multi-pair emission}
\label{sec:P4F}
The wave-vector for multi-mode SFWM squeezed state can be expressed in terms of the Fock bases as \cite{Christ2011}, 
\begin{eqnarray}
|\Psi\rangle &= \underset{k}{\otimes} \enskip \sqrt{1-x_k}~\sum_{n_k=0}^\infty ~\sqrt{x^{n_k}}~ |n_k,n_k\rangle  \label{eq:multimode}
\end{eqnarray}
Where $k$ represents the $k^{th}$ optical mode, $x_k$ is the squeezing strength of $k_{th}$ mode and $|n_k,n_k\rangle$ represents $n$ number of signal and idler photons in $k_{th}$ mode. If the density matrix from each source is $\hat{\rho} = |\Psi\rangle \langle \Psi|$, the lumped detection efficiency of a idler photon is $\eta_i$, and the probability of detecting $k$ idler photon is $P_{i}(k)$ for a generic non photon-number-resolving detector\cite{Kok2000}, then the reduced density matrix of the heralded signal photons for multi-mode twin-beam squeezer can be expressed by detecting at least one idler photon in at least one of the Schmidt modes $k$ with normalization constant $N$,
\begin{equation}
\hat{\rho}_s = N \sum_{\substack{n_1,...n_k=0\\n_1+...n_k\geq1}}^\infty P_i(n_1+...n_k) \langle n_1|...\langle n_k |\hat{\rho}| n_k \rangle...| n_1 \rangle
\end{equation}
In principle, photons propagating through a photonic circuit can be described as a general interferometer expressed as an unitary transformation matrix $U$. Considering the input optical modes of $U$ from source 1 and source 2 are $\hat{a}_{1i}$ and $\hat{a}_{2j}$ respectively with $i$ and $j$ as $i^{th}$ and $j^{th}$ Schmidt modes, and the output optical modes of the transformation are $\hat{c}^\dagger_{ij}$ and $\hat{d}^\dagger_{ij}$, then the transformation for distinguishable photons will be,
\begin{eqnarray}
&\hat{a}^\dagger_{11} \rightarrow U_{11}\hat{c}^\dagger_{11} + U_{12}\hat{d}^\dagger_{11};   \qquad&  \hat{a}^\dagger_{12} \rightarrow U_{11}\hat{c}^\dagger_{12} + U_{12}\hat{d}^\dagger_{12} \label{eq:xformation1}\\
&\hat{a}^\dagger_{21} \rightarrow U_{21}\hat{c}^\dagger_{21} + U_{22}\hat{d}^\dagger_{21};  \qquad&  \hat{a}^\dagger_{22} \rightarrow U_{21}\hat{c}^\dagger_{22} + U_{22}\hat{d}^\dagger_{22}\label{eq:xformation2}
\end{eqnarray}
For indistinguishable case, we can assume that both first and second Schmidt mode of source 2 will be the same as source 1 in the above equations: $\hat{c}^\dagger_{21}\rightarrow\hat{c}^\dagger_{11}$ and $\hat{c}^\dagger_{22}\rightarrow\hat{c}^\dagger_{12}$ and same treatment for mode $\hat{d}^\dagger$. Therefore, the probability of detecting two heralding idler photons, and heralded signal photons at both of the output modes of the unitary can be recorded as a four-fold coincidence event $P_{4F}$ describing heralded two-photon quantum interference. Thus, after the unitary transformation, $\hat{\rho} \xrightarrow{U} \hat{\rho'}$, the probability of detecting a four-fold event for the single mode squeezed state will be,
\begin{align}
P_\mathrm{4f} =  N_1 N_2 \sum_{\substack{n_{c_1},...n_{c_k}=0\\n_{c_1}+...n_{c_k}\geq1}}^\infty \sum_{\substack{n_{d_1},...n_{d_k}=0\\n_{d_1}+...n_{d_k}\geq1}}^\infty  & P_s(n_{c_1}+...n_{c_k})P_s(n_{d_1}+...n_{d_k}) \nonumber\\
& \langle n_{c_1}|\langle n_{d_1}|...\langle n_{c_k}|\langle n_{d_k}| \hat{\rho'} |n_{c_1}\rangle |n_{d_1}\rangle...|n_{c_k}\rangle|n_{d_k}\rangle \label{eq:4F_nov}
\end{align}
where $P_s(k)$ is the probability of detecting $k$ signal photons considering a lump detection efficiency $\eta_s$, and $c_k$ and $d_k$ are output optical modes of the $U$. The equation remains similar for multi-mode squeezed state but with more modes in the outputs. In an MZI, the phase $\Phi_{MZI}$ is scanned to get the minimum and maximum four-fold coincidences which is used to estimate visibility and infer indistinguishability. Typically, the visibility of a MZI fringe \cite{PhysRevLett.65.1348} is defined by,
\begin{equation}
%\label{eqn:Vmzi}
V_\mathrm{MZI} = \frac{(P_\mathrm{4f})_\mathrm{max} - (P_\mathrm{4f})_\mathrm{min}}{(P_\mathrm{4f})_\mathrm{max} + (P_\mathrm{4f})_\mathrm{min}}.\nonumber
\end{equation}
Using Eq.~(\ref{eqn:Pind}) and Eq.~(\ref{eqn:Pdist}) we see that the ideal fringe visibility for indistinguishable photons is $V^\mathrm{ind}_\mathrm{MZI}=1$, and reduces to $V^\mathrm{dist}_\mathrm{MZI}=1/3$ for distinguishable single-photons. In Eq.~(\ref{eq:4F_nov}), adding a second Schmidt mode and setting multi-photon emission upto 10, we can numerically calculate an interference fringe as a function of MZI phase and device parameters such as the directional coupler splitting ratio $\eta_{DC}$. This equation has been used to fit the 4-fold data obtained from the measurements with the fitting parameters: maximum coincidence counts ($C_{max}$) and phase offset ($\phi_{off}$) which is due to inevitable slight path length mismatches between both arms of the MZI. For the data shown in Fig.~\ref{fig:fringes}(a), the fitting values are $C_{max}=242.7964$ and $\phi_{off} = -0.2383\pi$.
\begin{figure}[htbp]
\centering
\includegraphics[width= \linewidth]{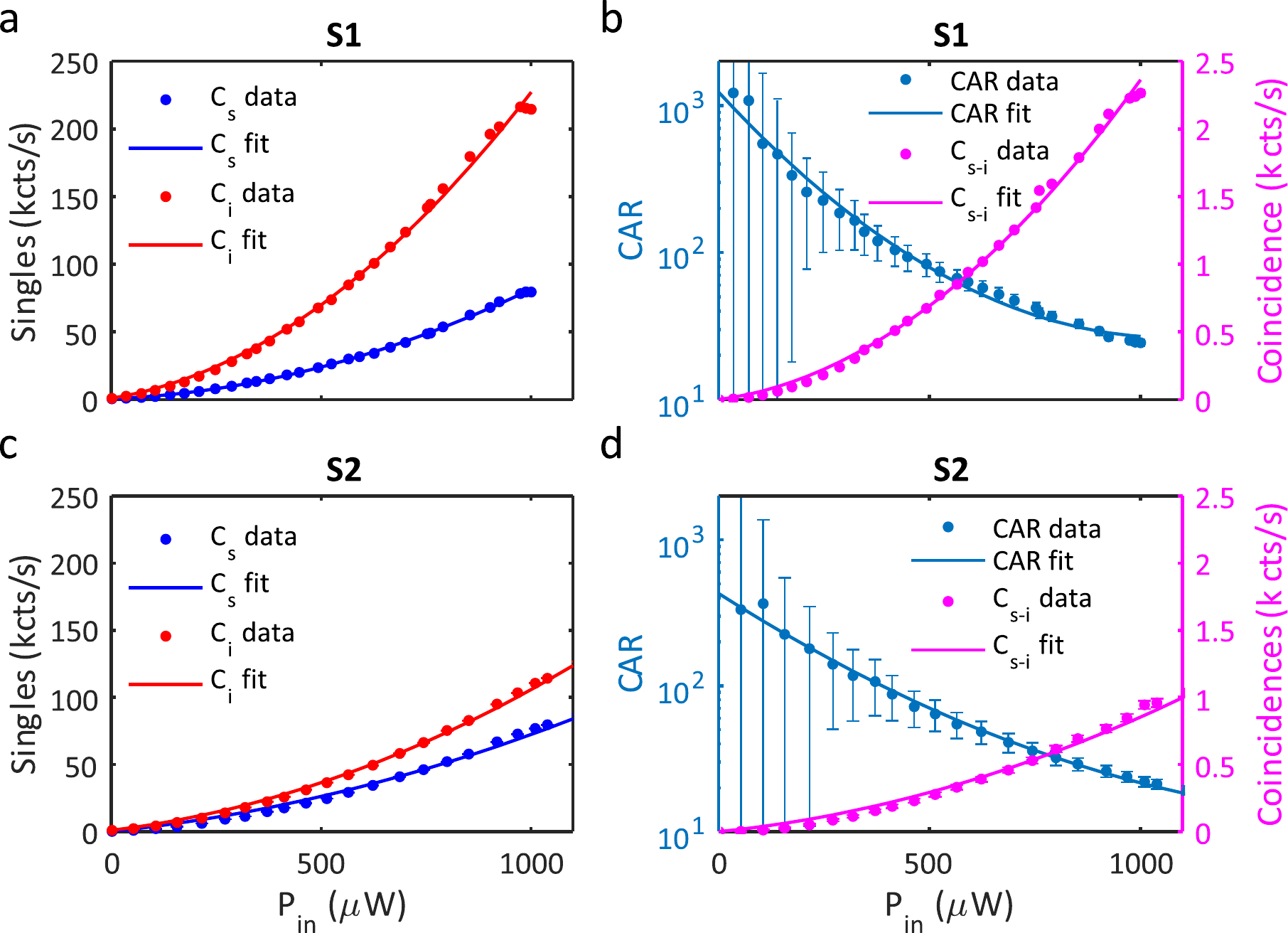}
\caption{\label{fig:nbar}Brightness for source S1 and S2. Signal and idler singles counts for S1 (a) and S2 (c). Signal-idler coincidence counts and the corresponding coincidence to accidental ratio (CAR) for S1 (b) and S2 (c).}
\end{figure}
%%%%%%%%%%%%%%%%%%%%%%%%%%%%%
\section*{Appendix C: Brightness ($\bar{n}$) estimation}\label{sec:nbar}
If the lumped efficiency of the four wave mixing process is $\gamma_{eff}$ ($\mathrm{pairs/s/mW^2}$), the input power is $P_{in}$ and the signal and idler photons detection efficiencies are $\eta_s$ and $\eta_i$, then the rate of detecting only signal photons, only idler photons and signal-idler coincidences from each sources can be written as,
\begin{align}
&C(s) = (\eta_s\gamma_{eff}) P_{in}^2 + \beta_s P_{in} + DC_s\\
&C(i) = (\eta_i\gamma_{eff}) P_{in}^2 + \beta_i P_{in} + DC_i\\
&CC(s,i) = (\eta_i\eta_s\gamma_{eff}) P_{in}^2 + ACC\\
&ACC = C(s)C(i)\tau
\end{align}
Here, $\beta_s$ and $\beta_i$ are the linear noise photon terms (e.g. pump leakage, scattered light, broken pairs etc.) including the detection efficiency, $DC$ represents dark counts and $ACC$ represents accidentals where $\tau$ is the size of the integration window. The factor $\gamma_{eff}$ lumps the total SFWM strength of the resonator. Fitting the above equations with the experimental data, $\gamma_{eff}$ can be estimated. If the repetition rate of the laser is $R$, then for an input power $P_{in}$, the average photon number generated per pulse will be, $\bar{n} = \gamma_{eff} P_{in}^2/R$. Our sources S1 and S2 have brightness values $\bar{n}_1 = 0.100\pm0.012$ and $\bar{n}_2 = 0.122\pm0.012$ with 95\% confidence interval as in Table~\ref{tab:2F}. Therefore, $\bar{n} = \sqrt{n_1n_2} = 0.110\pm0.012$. The extra losses in the signal and idler detection channels are due to the routing from the experimental setup to the single-photon detection systems. The average transmission per channel is $\eta=1.36\%$.
\begin{table}[ht]
\centering
\caption{\label{tab:2F} Average photon number per pulse estimation}
\begin{tabular}{l|l|l|l|l}
\hline
Source & $\eta_s$ & $\eta_i$ & $\gamma_{eff}$ & $\bar{n}$ @ $1~mW$ \\
\hline
1 & $0.0080$ & $0.0135$ & $5.013$ & $0.100 \pm 0.012$ \\
\hline
2 & $0.0111$ & $0.0287$ & $6.130$ & $0.122 \pm 0.012$ \\
\hline
\end{tabular}
\end{table}
The values of the $\beta_s$ = \{$12.9979$, $32.2330$\} $\mathrm{kcounts/s/mW}$, and $\beta_i$ = \{$49.9532$, $37.3080$\} $\mathrm{kcounts/s/mW}$ respectively for the two sources.

\bibliographystyle{unsrtnat}

%\bibliography{References}

\end{document}